\documentclass[9pt,twocolumn,twoside]{osajnl}

\journal{jocn} 

\setboolean{shortarticle}{false}
\usepackage{multirow}
\usepackage{enumitem}
\usepackage{comment}

\title{Semi-Automatic Line-System Provisioning with Integrated Physical-Parameter-Aware Methodology: Field Verification and Operational Feasibility}

\author[1,*]{Hideki Nishizawa}
\author[2]{Giacomo Borraccini}
\author[1]{Takeo Sasai}
\author[2]{Yue-Kai Huang}
\author[1]{Toru Mano}
\author[1]{Kazuya Anazawa}
\author[3]{Masatoshi Namiki}
\author[3]{Soichiroh Usui}
\author[1]{Tatsuya Matsumura}
\author[1]{Yoshiaki Sone}
\author[4]{Zehao Wang}
\author[1]{Seiji Okamoto}
\author[1]{Takeru Inoue}
\author[2]{Ezra Ip}
\author[2]{Andrea D'Amico}
\author[4]{Tingjun Chen}
\author[5]{Vittorio Curri}
\author[2]{Ting Wang}
\author[6]{Koji Asahi}
\author[1]{Koichi Takasugi}

\affil[1]{NTT Network Innovation Labs, 1-1 Hikarinooka, Yokosuka, Kanagawa, Japan}
\affil[2]{Optical Networking \& Sensing Department, NEC Laboratories America Inc., 4 Independence Way, Suite 200, Princeton, NJ 08540, USA}
\affil[3]{NTT Network Innovation Center, 3-9-11 Midori-cho,Musashino-shi,Tokyo, Japan}
\affil[4]{Department of Electrical and Computer Engineering, Duke University, Durham, North Carolina 27708, USA}
\affil[5]{Department of Electronics and Telecommunications, Politecnico di Torino, Corso Duca degli Abruzzi 24, Torino (TO), 10129, Italy}
\affil[6]{NEC Corporation, 1131, Hinode, Abiko, Chiba, Japan}

\affil[*]{hideki.nishizawa@ntt.com}

\begin{abstract}
We propose methods and an architecture to conduct measurements and optimize newly installed optical fiber line systems semi-automatically using integrated physics-aware technologies in a data center interconnection (DCI) transmission scenario. 
We demonstrate, for the first time, digital longitudinal monitoring (DLM) and optical line system (OLS) physical parameter calibration working together in real-time to extract physical link parameters for transmission performance optimization.
Our methodology has the following advantages over traditional design: a minimized footprint at user sites, accurate estimation of the necessary optical network characteristics via complementary telemetry technologies, and the capability to conduct all operation work remotely. The last feature is crucial, as it enables remote operation to implement network design settings for immediate response to quality of transmission (QoT) degradation and reversion in the case of unforeseen problems.
We successfully performed semi-automatic line system provisioning over field fiber networks facilities at Duke University, Durham, NC. The tasks of parameter retrieval, equipment setting optimization, and system setup/provisioning were completed within 1 hour. The field operation was supervised by on-duty personnel who could access the system remotely from different time zones. By comparing Q-factor estimates calculated from the extracted link parameters with measured results from 400G transceivers, we confirmed that our methodology has a reduction in the QoT prediction errors ($\pm 0.3$ dB) over existing design ($\pm 0.6$ dB).
\end{abstract}

\setboolean{displaycopyright}{false} 

\begin{document}

\maketitle

\section{Introduction}


The implementation of optical networks continues to expand all over the world as the rapid adoption of cloud, AI/ML, big data, analytics, and AR/VR technologies boosts the number of data center interconnection (DCI) installations.
In recent years, use cases for new applications such as interactive live music entertainment and remote operation have been examined across industries, and the network latency requirements have been clarified~\cite{iowngf}.
There is also high demand for low-latency DCI for computer-to-computer communication, and new optical fiber routes with shorter distances are being installed to meet this demand even when the same route already exist but with a longer distances.
In addition to the need for latency, fast DCI provisioning will be required as data centers become more distributed because many data center operators have adopted zero-touch provisioning technology for devices such as servers and network switches.
Against this background, the expectations for automation of optical network operations are growing.
In this context, the optical line system (OLS)~\cite{TIP2021} plays a fundamental role in the network architecture, being a critical part in the automation of the management infrastructure.

To accelerate such automation in optical network design, various organizations are working toward the disaggregation of optical transmission systems~\cite{tip_oopt, openroadm} and the development of open optical network optimization tools such as GNPy~\cite{ferrari2020gnpy} and Mininet-Optical~\cite{Mininet-Optical}.
These tools are based on Gaussian noise (GN) models and enable operators to analytically estimate the quality of transmission (QoT) with limited computational resources by inputting the optical network system's characteristic parameters, such as the fiber type and erbium-doped fiber amplifier (EDFA) gain.
Furthermore, over the last decade, automatic optimization methods for optical network operation via machine learning have been reported by both academia and industry~\cite{lu2021performance} because machine learning techniques have achieved phenomenal success in revolutionizing many science and engineering disciplines.

\subsection{Issues and Our Approach}

Despite years of rigorous research, however, automatic operation has not yet gained broad acceptance in commercial fiber-optic networks.
Khan presented issues for the commercial use of machine learning technology~\cite{khan2022machine}.
By combining the issues that he identified with those that we are cognizant of, we can summarize the issues with automatic operation of optical networks via machine learning technology as follows.

\begin{itemize}
    \item (i) Insufficient verification: Much of the research to date has only verified the capabilities of individual technologies separately in an ideal lab environment.
    In addition, those technologies only support a subset of the optical system’s characteristic parameters that are needed for automatic operation.

    \item (i) Adaptability to conventional methods: The existing trained workforce and established technical practices are adapted to conventional non-automatic methods.
    Scenarios for migration from existing operations need to be presented to implement new technologies/methods.

    \item (i) Transparency and explainability: Operators must guarantee their network service quality according to service level agreements with users.
    When a failure occurs, in some cases the operator must identify the failure's root cause and location and report them to the user.

    \item (ii) Cost constraints: The cost of newly installed equipment, use of an extended set of tools, and additional workforce for automatic operation must be compensated for by benefits that outweigh them.

    \item (iii) Generalizability: If methods or models are sensitive to changes in data or in a device type/vendor because of temporal variations or link modifications, then they cannot be applied in dynamic, diverse, wide-area network scenarios.
\end{itemize}

To address these issues, we apply the following solution approaches in this paper.
The notation (i), (ii) and (iii) above and below indicate the relationships between the issues and solutions.

\begin{itemize}
    \item (i) Integrated physical parameter estimation methods: Perform link parameter estimation and refinement for automatic provisioning based on physical models, not a black-box approach, including digital longitudinal monitoring (DLM) and OLS physical parameter calibration method.

    \item (i) Semi-automatic operation: Work that involves decisions, such as margin design and identification of fault locations, should be done by humans with visualized span-by-span physical parameters and bit error rate (BER) analysis.

    \item (i) Field verification: Operational feasibility should be confirmed with field fiber, including underground and aerial segments.

    \item (ii) CAPEX suppression: Small footprints should be maintained at user sites by using data from digital coherent receivers as a by-product of digital signal proccessing (DSP) algorithms, such as DLM, and by sharing additional equipment at carrier edges among many users.

    \item (ii) OPEX reduction: Reduced on-site work, remote operation from different time zones, and reduced construction time should all be implemented.

    \item (iii) Hybrid controller: Acquired/analyzed data for automatic operation should be classified and stored locally or remotely according to its volume/latency requirements.

    \item (iii) Open approach: By using open source software (OSS) for analytical model and multi-vendor device operation, the same analysis and device control can be performed anywhere in the world like at the local site.
\end{itemize}

In this paper, we assume a data center exchange (DCX) service as a use case where automated operations will be introduced first.
DCX services directly connect any data centers distributed in a metro area via optical fiber to quickly establish high-capacity optical paths from each company’s cloud services to mobile edge computing device and the users connected to them~\cite{nishizawa2024fast}.

The remainder of this work is organized as follows.
After introducing the related works in Sec.~\ref{sec:related_works} to depict the state of the art, 
Sec.~\ref{sec:methodology} shows the proposed physical-parameter-aware methodology to semi-automatically perform the provisioning of an optical line and the description of how the QoT measurements are carried out and their comparison with the prediction.
Sec.~\ref{sec:control_architecture} illustrates the transmission scenario and the controller architecture implemented to manage the provisioning operations.
After describing the experimental setup in Sec.~\ref{sec:experimental_setup}, we report the achieved results of 1-hour provisioning and maintenance, including execution time, physical parameter visualization, performance validation, baseline comparison, system Q-factor fluctuation, and operational considerations in Sec.~\ref{sec:results}.
Sec.~\ref{sec:future_challenges} discusses future challenges for performance improvement and the applicability of the technique to new use cases, and Sec.~\ref{sec:conclusion} concludes the paper.

\begin{figure*}[!t]
\centerline{\includegraphics[width=0.9\linewidth]{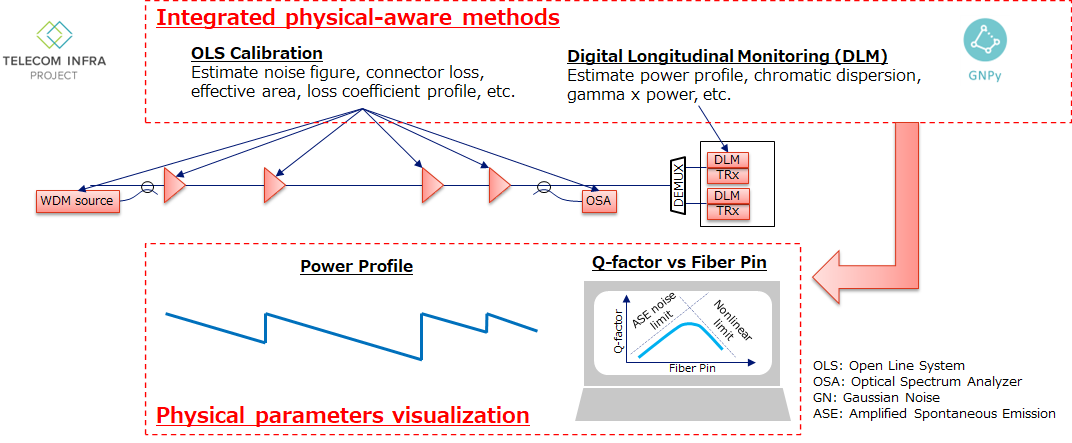}}
\caption{Key enablers: Physical parameters visualization with integrated physical-aware methods.}
\label{fig:key_enablers}
\end{figure*}

\section{Related Works} \label{sec:related_works}

As described in section 1.A, Khan summarized the issues for the commercial use of machine learning technology~\cite{khan2022machine}.
That paper briefly mentioned the use of rich monitoring data readily available in a digital coherent receiver as a by-product of DSP algorithms such as DLM, instead of reliance on data from standalone optical performance monitor (OPM) devices.
However, the paper did not give an architecture or experimental results.
Another paper~\cite{nishizawa2023dynamic} published around the same time presented a similar idea with experimental results.
That paper aimed to automatically connect customer-owned transceivers via alien access links (AALs) with a maximum data rate.
It proposed an architecture and protocol for cooperative optical path design between a customer and carrier, by using DLM to estimate link parameters.
As for studies of DLM alone, a number of experiments were described in~\cite{sasai2022digital}.
The model parameters such as the losses, CD, gains, and filter responses were estimated, and a DLM was then applied to visualize the multi-span characteristics of the fibers, amplifiers, and filters.
However, ~\cite{nishizawa2023dynamic, sasai2022digital} did no include field trial results or discuss the short execution time (an offline DLM system was used).
Previously, an optical line physical parameter calibration (OL calibration) method was proposed to improve QoT estimation by calibrating an optical link's physical model parameters after installation~\cite{borraccini2024optical}.
Although the accuracy of optical signal-to-noise ratio (OSNR) estimation results was presented, no BER analysis with estimated generalized signal-to-noise ratio (GSNR) results was conducted to verify the QoT prediction and optimization.
As for BER analysis based on GN approximation, Kaeval et al. proposed an approach to estimate the End-to-end (EtE) GSNR from the estimated GSNR for each link constituting the EtE route~\cite{kaeval2022concatenated}.
Our group validated a transceiver BER-OSNR model and applied Q-factor estimation for short-reach systems~\cite{mano2023modeling}.
We also reported verification results from field trials~\cite{mano2023modeling, nishizawa2024fast}.
However, these papers involving BER analysis only described transceiver provisioning methods for optical coherent mode selection, rather than new OLS provisioning with EDFA configuration.

\section{Methodology} \label{sec:methodology}

Fig.~\ref{fig:key_enablers} provides on overview of the key enablers of the proposed solution approach.
Our methodology integrates two major physics-aware techniques for link parameter estimation, which are performed during the optical line provisioning in the following order: the DLM and the OLS calibration.
Each technique is carried out in the following two different steps, which are the data collection and the physical parameter extraction.

\subsection{Digital Longitudinal Monitoring}

DLM is a transponder-based monitoring technique that visualizes physical characteristics of multiple link components along the fiber-longitudinal direction solely by processing received data-carrying signals~\cite{sasai2022digital}.
Among parameters estimated by DLM, longitudinal optical power is of particular importance since it enables the localization of multiple lumped losses in a link and the extraction of fiber parameters.
From its working principle, DLM estimates the product of fiber nonlinear coefficients and longitudinal optical power $\gamma(z)P(z)$~\cite{sasai2023performance}.
Also, with a priori information of fiber-span lengths, span-wise chromatic dispersion (CD) parameters can also be estimated by calibrating positions of EDFAs, indicated by longitudinal power profiles.

The first step of the methodology is the data collection and consequent parameter extraction using DLM.
The only DLM signal is transmitted along the EtE path, through both AALs and installed optical line within the carrier link.
The EDFAs are set in constant output power at a level allowing the DLM to reach a suitable accuracy and resolution (roughly 12~dBm).
In this paper, we particularly used the linear least squares algorithm for DLM as described in~\cite{sasai2024linear}, allowing for a rapid estimation of the absolute value of $\gamma(z)P(z)$.
Knowing the lengths of the fiber spans, $L_S$, the parameters extracted per each fiber span using the data collected through the DLM are: the loss coefficient, $\alpha$, at the DLM frequency, the CD, $D$, the lumped loss positions and values, $l(z)$, the product between the nonlinear coefficient, $\gamma$, and the fiber span input power, $P$.
The latter can be interpreted as the product between the power measured by the EDFA output photodiode and the input connector loss.
For this reason, the separation of $\gamma$ and $P$ is achieved in the next step during the OL calibration, since it jointly defines the entity of the input connector loss and the effective area of each fiber span.

\begin{figure*}[!t]
\centerline{\includegraphics[width=0.9\linewidth]{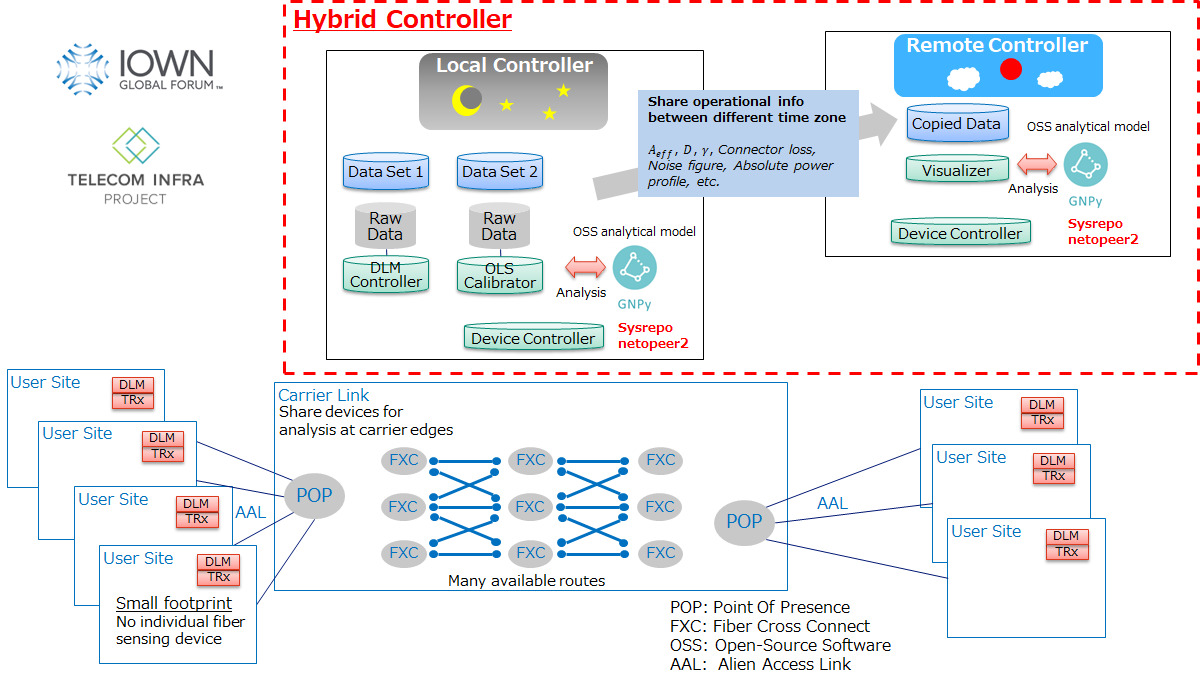}}
\caption{Key enabler: Hybrid controller in DCX architecture.}
\label{fig:hybrid_controller}
\end{figure*}

\subsection{Optical Line Physical Parameter Calibration}

Following the installation of an optical line, the second method is a telemetry-based technique improving QoT estimation by determining the physical parameters using only optical power monitors integrated in optical amplifiers and one optical spectrum analyzers (OSA) at both edge nodes of the considered optical line~\cite{borraccini2024optical}.
The retrieved physical parameters can then be applied in physics simulation model such as GNPy to define the optimal operation working point.

A full C-band WDM comb is introduced at the first edge node using the combination of an amplified spontaneous emission (ASE) source and a wavelength-selective switch (WSS), which can be readily available at the carrier's edge node equipped with ROADM.
The WSS is used to flatten the ASE source and to create arbitrary spectral holes to perform OSNR measurements.
The data collection is done by changing the gain and tilt settings of each EDFA in the carrier link, while measuring the corresponding spectrum at both the transmitting and receiving nodes and the input/output powers for each of the EDFAs.
%
%
Then, the parameter retrieved for each span are the loss coefficient function, $\alpha(f)$, the input, $l(z=0)$, and output, $l(z=L_S)$ connector losses, and the function of the average EDFA noise figure (assumed flat in frequency) with respect to the gain parameter, $\mathrm{NF}$.
The estimation of the overall gain ripple profile is based on the assumption that all EDFAs exhibit the same intrinsic gain ripple profile.
The effective area, $A_{eff}$, is indirectly derived from the nonlinear coefficient, $\gamma$, removing the estimated input connector loss from the value provided by the DLM.
The aim of the physical parameter extraction is to determine the set of physical parameters that minimizes the cost function associated with the averaged errors between measured and simulated profiles in terms of signal power and OSNR.

In this way, comprehensive characterization of fibers and amplifiers is enabled by combining DLM and OL calibration.
By inputting the physical parameters obtained by OL calibration and DLM into GNPy simulation model, the GSNR for link transmission for every WDM channel can be estimated by providing the channel input power. 
Assisted with transceiver BER vs optical signal-to-noise ratio (OSNR) characteristics ~\cite{mano2022accuracy, mano2023modeling}, we can then calculate the sum of impairments generated by the OLS and the transceivers to estimate Q-factor corresponding to the pre-Forward Error Correction (FEC) BER at receiver~\cite{nishizawa2024fast}.
These methods allow operators to visualize the information they need for network operation such as margin design (Physical parameters visualization) and performance optimization.
Thanks to the combined DLM and OLS calibration strategy, and analytical physical model using GN approximation, computation time is significantly reduced, and results can be displayed immediately to operators.

\subsection{QoT Estimation \& Measurement}

Following what was stated in~\cite{nishizawa2024fast}, below we define how the effective EtE SNR of the transmission channel is measured and how it is compared to the prediction of the physical model, in this case GNPy.

Given a specific modulation format, $\mathrm{MF}$, the measured pre-FEC BER can be modelled as a function, $\Psi_{\mathrm{MF}}$, of the overall SNR across an EtE optical path:
\begin{equation} \label{eq:ber_snr}
    \mathrm{BER}=\Psi_{\mathrm{MF}}(\mathrm{SNR})\,,
\end{equation}
\begin{align} \label{eq:snr}
    \mathrm{SNR}^{-1} &= \mathrm{SNR_{TRx}^{-1}} + \mathrm{SNR_{ASE}^{-1}} + \mathrm{SNR_{NLI}^{-1}} \\ \nonumber
                      &= \mathrm{SNR_{TRx}^{-1}} + \mathrm{GSNR^{-1}}\,,
\end{align}
where for the 16QAM:
\begin{equation} \label{eq:psi}
    \Psi_{\mathrm{MF}}=\frac{3}{8}\mathrm{erfc}\left(\sqrt{\frac{\mathrm{SNR}}{10}}\right)\,,
\end{equation}

From the back-to-back BER vs. OSNR measurement curve, the effective SNR of the transceiver, $\mathrm{SNR_{TRx}}$ can be derived by least square fitting method.
For estimating the EtE SNR, GNPy model first provides the estimation of the GSNR which combines the contributions related to the ASE noise and the NLI impairment in the optical line.
The predicted EtE SNR is achieved taking into account the effect of the transceiver SNR, following Eq.~\ref{eq:snr}.
For verifying the estimation accuracy using BER measurement, the SNR of the EtE connection is calculated by inverting Eq.~\ref{eq:ber_snr}:
\begin{equation}
    \mathrm{SNR}=\Psi_{\mathrm{MF}}^{-1}(\mathrm{BER})\,.
\end{equation}
We then subtract the estimated and measured SNRs by an arbitrary value to obtain the relative Q-factors for the channels under test, as to refrain from disclosing the exact performance of the transceiver related to $\mathrm{SNR_{TRx}}$.
\begin{figure}[!b]
\centerline{\includegraphics[width=\linewidth]{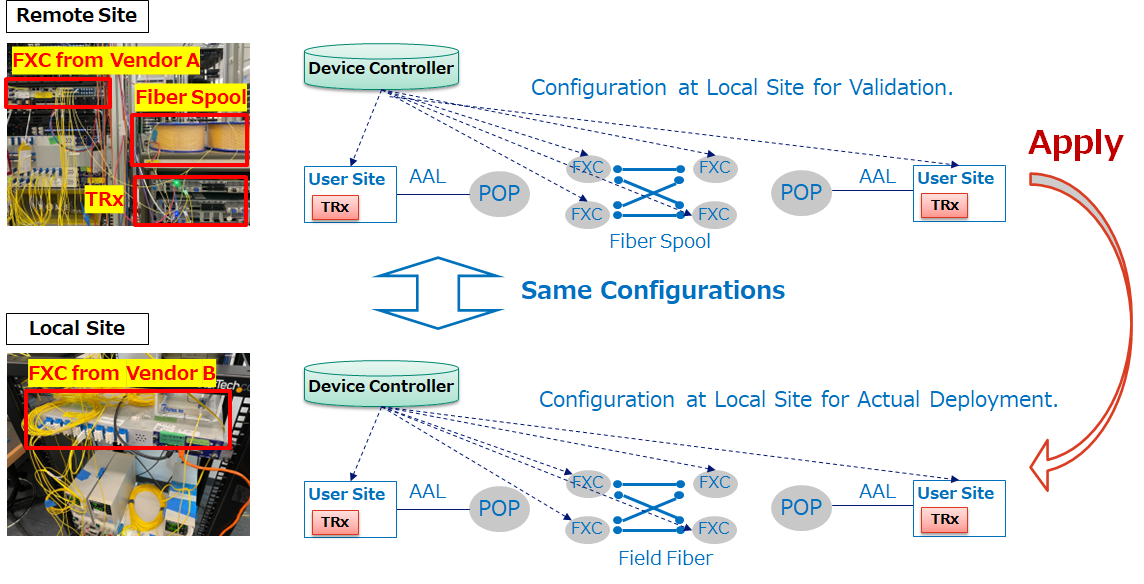}}
\caption{Validation and OLS installation by Device controller.}
\label{fig:device_controller}
\end{figure}

\section{Control Architecture} \label{sec:control_architecture}

Fig.~\ref{fig:hybrid_controller} shows one key enabler of the newly conceived DCX architecture: the hybrid controller.
User sites are connected by carrier links owned by a carrier, which provides a point of presence (POP) at carrier edges~\cite{nishizawa2024fast}.
User equipment can connect to the POP via AALs and directly to other user equipment without electrical conversion.
In this service, many users are connected to each other via carrier links where optical switches are deployed, so that analysis equipment deployed at the carrier edge can be shared among users.
In addition, DLM can be applied to user equipment to retrieve the physical parameters of fiber links without adding new monitoring devices on the user site.

The use of a hybrid controller (local/remote) architecture with OSS analytical model and common interface for multi-vendor device control enables operators to perform common operating procedures across different time zones without sharing big data.
This architecture's features are as follows: (1) the same analysis and device control can be performed anywhere in the world by using OSS for analysis models and multi-vendor device operation; and that (2) raw data with a large volume or large impact on the computation time is stored only at the local site.
Hybrid controller consists of four main components; DLM controller, OLS calibrator, Visualizer, and Device controller.

The DLM controller at Local site facilitates DLM functionality on the Muxponder, which retrieves raw waveform data from the 130-Gbaud-class real-time transceiver DSP. The physical parameters extracted by DLM are stored as data set 1 after processing.
Likewise, the OLS calibrator controls and conducts measurements using the WSS, optical amplifiers, and OSA. It then performs physical parameters extraction by processing the measured data and stored the parameters as data set 2.
We utilize GNPy at the local site for link transmission modelling and QoT performance optimization and visualization. 
As shown in Fig.~\ref{fig:provisioning_proc}, all physical parameters necessary for visualization are computed by DLM + OL calibration, utilizing data set 1 and data set 2.
These parameters are copied to the remote controller. Here, the physical parameters for operation were shared between controllers, but the raw data extracted from DSP for DLM and generated in the process of OLS calibration are stored at local controller (not only to minimize latency, but these technical data may also be handled under non-disclosure agreement under certain situations).
The operator at the remote site can design the OLS by entering these physical parameters into GNPy.

Finally, Device controller performs fiber cross connect (FXC) and TRx configurations for OLS installation.
Here, note that the unified control interfaces for TRx have already been defined and standardized in MSA such as OpenROADM~\cite{openroadm} while that of FXC has not. Since each vendor has a different control interface for FXC, a common interface is essential to control local devices from a remote controller as in this case.
So, we utilized our defined unified control interface for FXC in~\cite{anazawa2024first}, and extend our controller developed in~\cite{anazawa2024first} for OLS installation.
Using them, the operator at the remote site can validate the OLS configurations using multi-vendor FXCs and TRx at the local site as shown in Fig.~\ref{fig:device_controller}.
According to the validation result, the operator makes the decision to start the service with newly installed OLS.
We note that our unified control interface and Device controller utilize OSS of Yang-based datastore Sysrepo~\cite{sysrepo} and NETCONF server netopeer2~\cite{netopeer2}, so the controller can be used at anywhere.

\section{Experimental Setup} \label{sec:experimental_setup}

\begin{figure}[!t]
\centerline{\includegraphics[width=\linewidth]{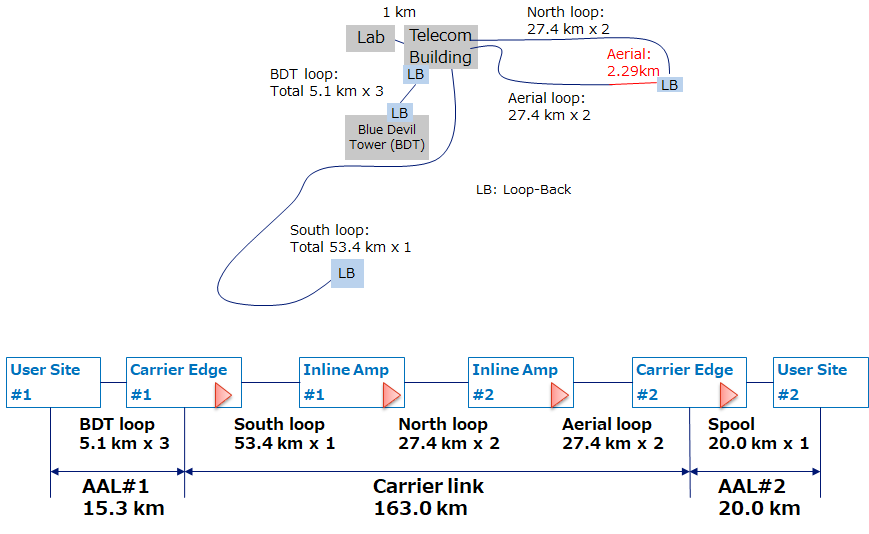}}
\caption{Fiber configuration.}
\label{fig:fiber_config}
\end{figure}


Fig.~\ref{fig:fiber_config} shows the link fiber routes and configuration that we used in our field trial.
All the field fibers (North, Aerial, Blue Devil Tower, and South loops) are single-mode fibers that start and end at the Duke University's Telecom Building, which is a short hop from the optical/wireless bed housing our test system.
All the fiber routes are entirely underground except for the Aerial loop, which includes a 2.29-km-long aerial section.
We used loop-backs (blue box) on the Blue Devil Tower (BDT) loop and South loops.
An EtE fiber route for the DCX network was constructed using field fiber and lab fiber spools.
AAL\#1 was 15.3-km long and comprised three 5.1-km fiber loops (the BDT loop, three times). The carrier link was 163.0 km long and comprised the South loop, the North loop (twice, connected back-to-back), and the Aerial loop (twice, connected back-to-back).
A 20 km spool fiber was used for AAL\#2. 

\begin{figure*}[!t]
\centerline{\includegraphics[width=\linewidth]{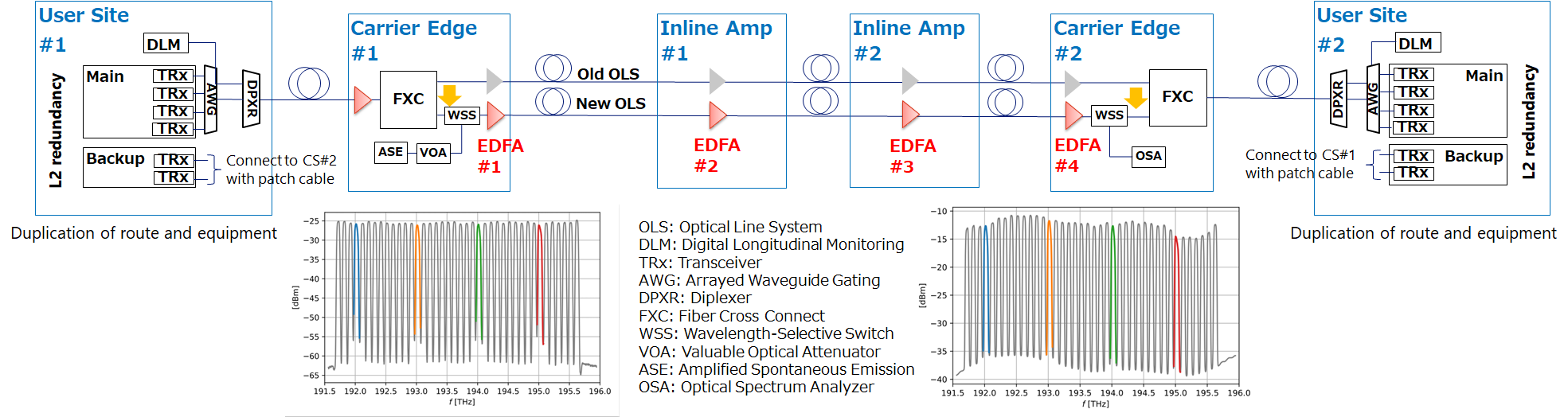}}
\caption{Network configuration.}
\label{fig:network_config}
\end{figure*}

Next, Fig.~\ref{fig:network_config} shows the network configuration in this experiment.
We assumed that DCX users applied L2 redundancy control with a duplicated route and muxponder, and no L1 redundancy control.
The back-up muxponders at User Sites \#1 and \#2 were connected with patch cable to emulate a backup link until the old OLS would be switched with the newly installed OLS.
The main muxponders had four 1-THz frequency spaced TRxs across the C-band. The four frequencies are multiplexed with the arrayed waveguide grating (AWG) with fixed 100-GHz spacing.
We deployed white-box muxponders that complied with TIP’s Phoenix requirements~\cite{tip_phoenix}, and we installed NEC’s NOS, which is based on the TIP Goldstone NOS~\cite{tip_goldstone}.
These muxponders used OpenROADM-compliant 400G CFP2-DCO pluggable transceivers from Fujitsu Optical Components and Lumentum.
We deployed a Fujitsu T950 muxponder for the DLM demonstration in this experiment, and the DLM signal was multiplexed with diplexer (DPXR) which combines it with the four 400G channels, as the DLM channel bandwidth is too wide to fit through the AWG. The DLM function will be implemented with TRx for small footprint in future.
Carrier Edges \#1 and \#2 had FXCs to switch the optical path route from the old OLS to the new OLS.
Carrier Edge \#1, which was the OLS' starting point, had an ASE source and a WSS to generate the full C-band DWDM transmission spectrum on the carrier link, with a total of 40 channels including DLM and four 400G channels at 100-GHz spacing (The DLM channel occupies 200-GHz slot).
Carrier Edge \#2, the end point, had a WSS and an OSA to measure the spectrum.
All EDFAs, in the Carrier Edges and in two In-line Amps (\#1 and \#2), were dual-stage with automatic gain control (AGC).

\section{Results: 1-Hour Provisioning \& Maintenance} \label{sec:results}

We introduce the following mathematical notations to allow easy reading of the results.
When considering an arbitrary frequency-dependent profile $p(f)$, $\overline{p(f)}$ denotes its average value across frequencies.
The error operator $\Delta\left[p(f)\right]$ quantifies the difference between measured and simulated metric values, and $\sigma_{p(f)}$ denotes the standard deviation over the frequency spectrum.

\subsection{Execution Time}

\begin{figure}[!b]
\centerline{\includegraphics[width=0.9\linewidth]{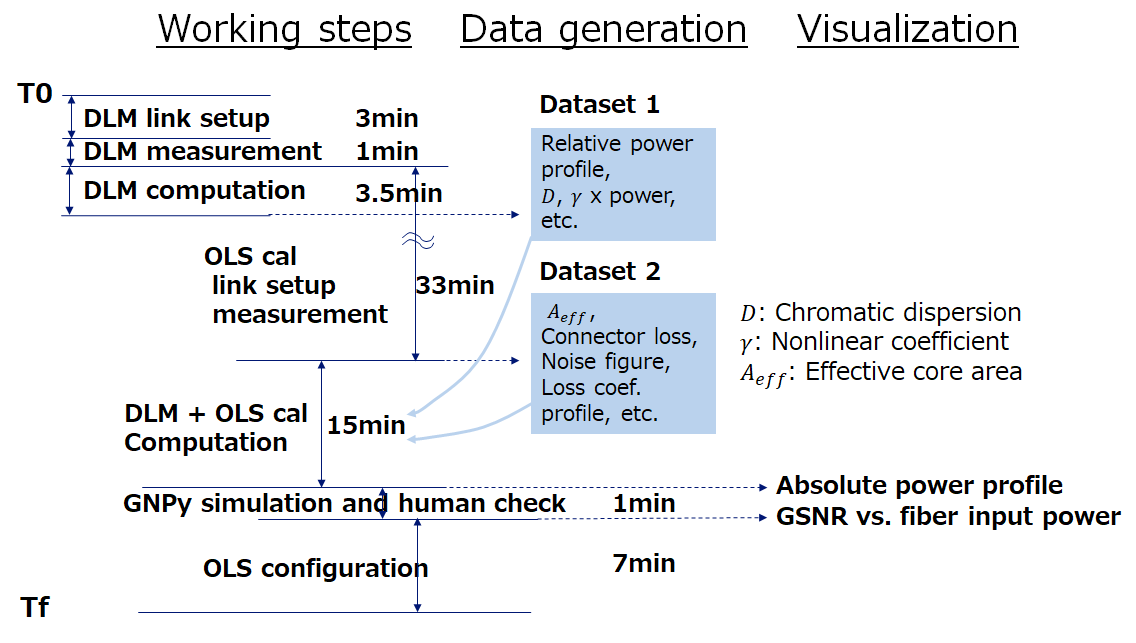}}
\caption{Provisioning procedure and execution time.}
\label{fig:provisioning_proc}
\end{figure}

\begin{figure}[!b]
\centerline{\includegraphics[width=\linewidth]{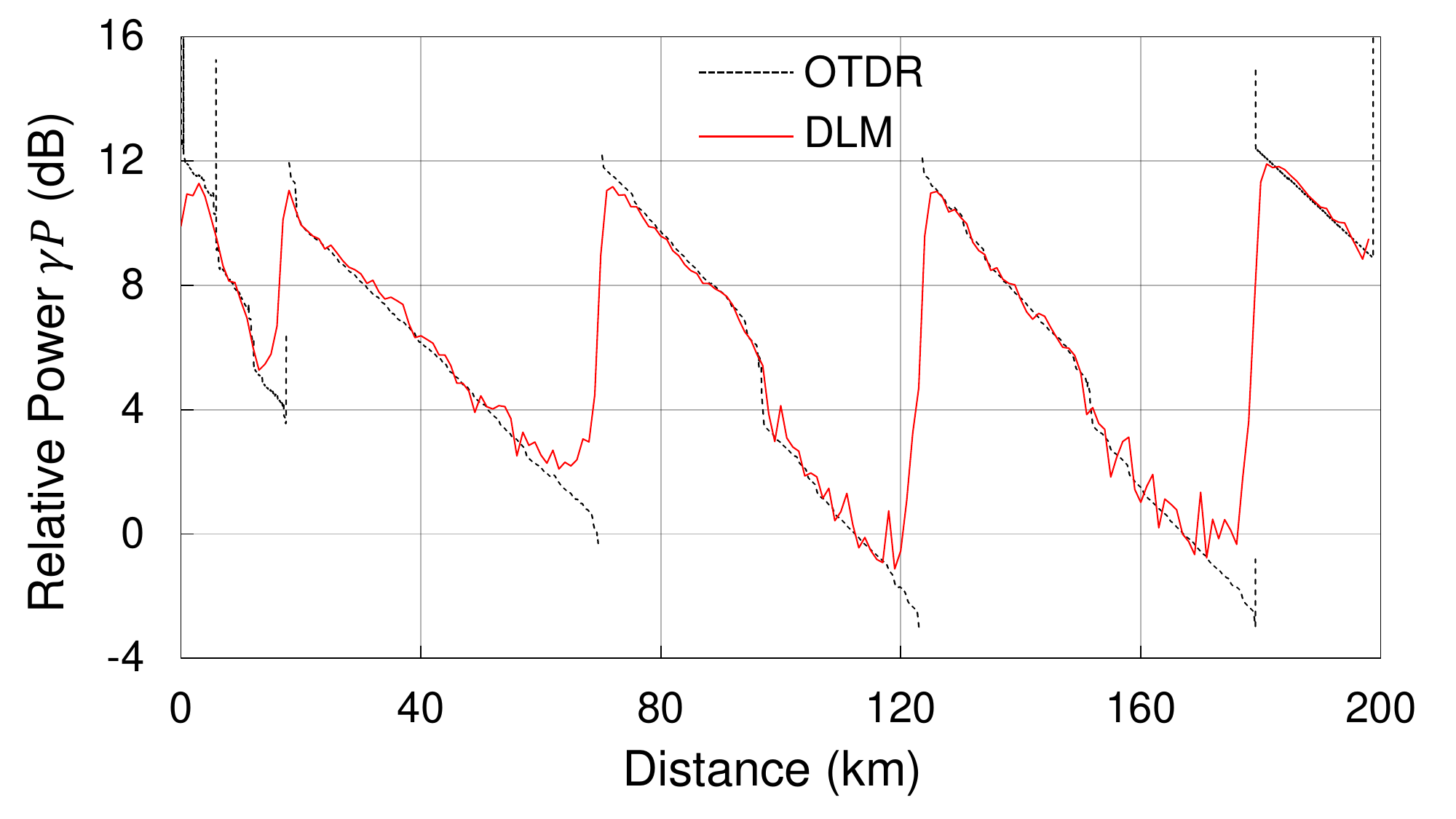}}
\caption{Relative power profile estimated by DLM.}
\label{fig:power_profile}
\end{figure}

\begin{table*}[!t]
\centering
\begin{tabular}{|c|c|c|c|c|c|ccc|}
\hline
\multirow{2}{*}{Span ID} &
  \multirow{2}{*}{\begin{tabular}[c]{@{}c@{}}$L_S$\\$[$km$]$\end{tabular}} &
  \multirow{2}{*}{\begin{tabular}[c]{@{}c@{}}$A_{eff}$\\$[\mu$m$^2$$]$\end{tabular}} &
  \multirow{2}{*}{\begin{tabular}[c]{@{}c@{}}$D$\\$[$ps/nm/km$]$\end{tabular}} &
  \multirow{2}{*}{\begin{tabular}[c]{@{}c@{}}$\gamma$\\$[$1/W/km$]$\end{tabular}} &
  \multirow{2}{*}{\begin{tabular}[c]{@{}c@{}}Total Loss\\$[$dB$]$\end{tabular}} &
  \multicolumn{3}{c|}{$l(z)$ [dB]} \\ \cline{7-9} 
  & &       &       &       &      & \multicolumn{1}{c|}{$z=0$} & \multicolumn{1}{c|}{$0<z<L_S$} & $z=L_S$ \\ \hline
AAL 1 & 17.6 & 82.2 & 17.94 & 1.28 & 11.7 & \multicolumn{1}{c|}{2.7}  & \multicolumn{1}{c|}{--}      & --    \\ \hline
CL 1 & 51.86 & 90.84 & 17.97 & 1.16 & 15.7 & \multicolumn{1}{c|}{3.33}  & \multicolumn{1}{c|}{2.25}      & 0.93    \\ \hline
CL 2 & 54.75 & 91.92 & 17.35 & 1.15 & 16.8 & \multicolumn{1}{c|}{2.73}  & \multicolumn{1}{c|}{2.40}      & 0.89    \\ \hline
CL 3 & 54.75 & 95.67 & 17.67 & 1.10 & 16.4 & \multicolumn{1}{c|}{2.57}  & \multicolumn{1}{c|}{1.28}      & 0.62    \\ \hline
AAL 2 & 20.0 & 83.0 & 16.71 & 1.27 & 5.0 & \multicolumn{1}{c|}{1.7}  & \multicolumn{1}{c|}{--}      & --    \\ \hline
\end{tabular}
\caption{Set of fiber physical parameters of the field trial. The physical parameters related to the carrier link are extracted using the proposed calibration methodology. The AAL physical parameters are derived combining the DLM technique and EDFA telemetry.}
\label{tab:fiber_info}
\end{table*}

Fig.~\ref{fig:provisioning_proc} shows provisioning procedure and execution time.
It includes not only the working steps but also the processes of data generation and visualization and their interrelationships.
T0 and Tf indicate the start and end times, respectively.
First, the fiber parameters were obtained using DLM.
The link set up took 3 minutes, including startup of the coherent module, and 1 minute for the measurement.
After the DLM measurement, the DLM computation and OL calibration setup and measurement ran in parallel, and the two processes took 3.5 minutes and 33 minutes, respectively. 
The DLM computation will produce the dataset 1, where the measurement results of OLS calibration were passed to the OLS calibration computation step to extract the  link parameters and produce dataset 2. 
Then, dataset 1 and dataset 2 were combined as input to GNPy model computation to visualize the link characteristics including the absolute power profile and relative Q-factors vs channel input power.
It took 15 minutes after the OLS calibration measurement to produce the visualization results, which were mostly utilized by the OLS calibration computation as GNPy computation time is negligible. 
The remote operator was given 1-minute to check the visuals provided by the physical-parameter-awared model, before making the decision to either adopt the new OL configuration or revert to the old configuration. The final OL configuration will take 7-minutes to complete, hence complete the entire task under one hour.
%
%

\subsection{Physical Parameter Visualization}

\begin{figure}[!t]
\centerline{\includegraphics[width=\linewidth]{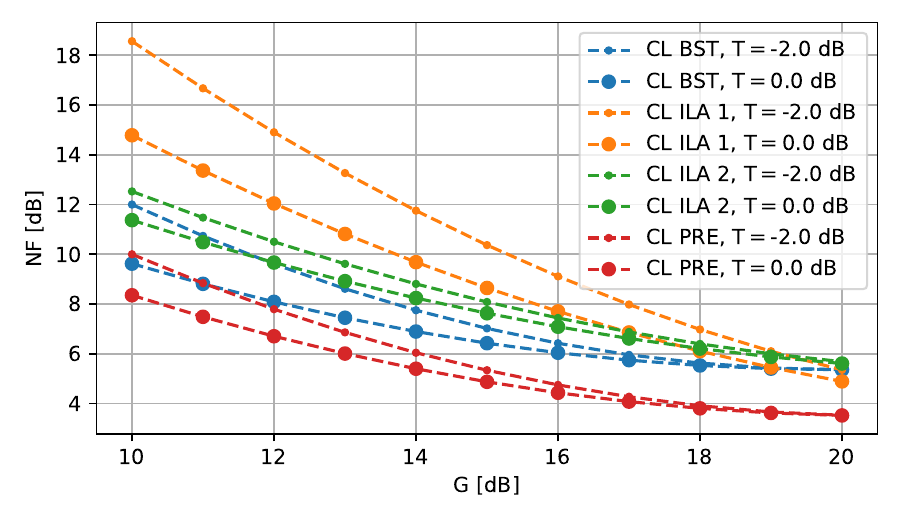}}
\caption{EDFA noise figure vs. target gain/tilt curves in the carrier link retrieved using the proposed methodology.}
\label{fig:edfa_params}
\end{figure}

Fig.~\ref{fig:power_profile} shows a relative power profile visualized by the DLM with the fiber configuration shown in Fig.~\ref{fig:fiber_config}.
The vertical axis and the horizontal axis indicate relative optical power ($\gamma P$) in dB and distance, respectively. DLM agrees well with optical time domain reflectometer (OTDR) traces and several lumped losses in field fibers were observed. The RMS error between OTDR traces and DLM was 0.45 dB (Amplifiers' positions +- 3 km were excluded from the RMS error calculation).
Tab.~\ref{tab:fiber_info} shows the set of physical parameters of each carrier link fiber span retrieved using the proposed methodology.
The fiber physical parameters of the AALs are retrieved using the DLM technique combined with the EDFA power monitor measurements.
Furthermore, Fig.~\ref{fig:edfa_params} shows EDFA noise figure versus target gain/tilt curves in the carrier link.
The average noise figure value for EDFAs in both ALLs is assumed 6.5~dB.

\begin{figure}[!b]
\centerline{\includegraphics[width=\linewidth]{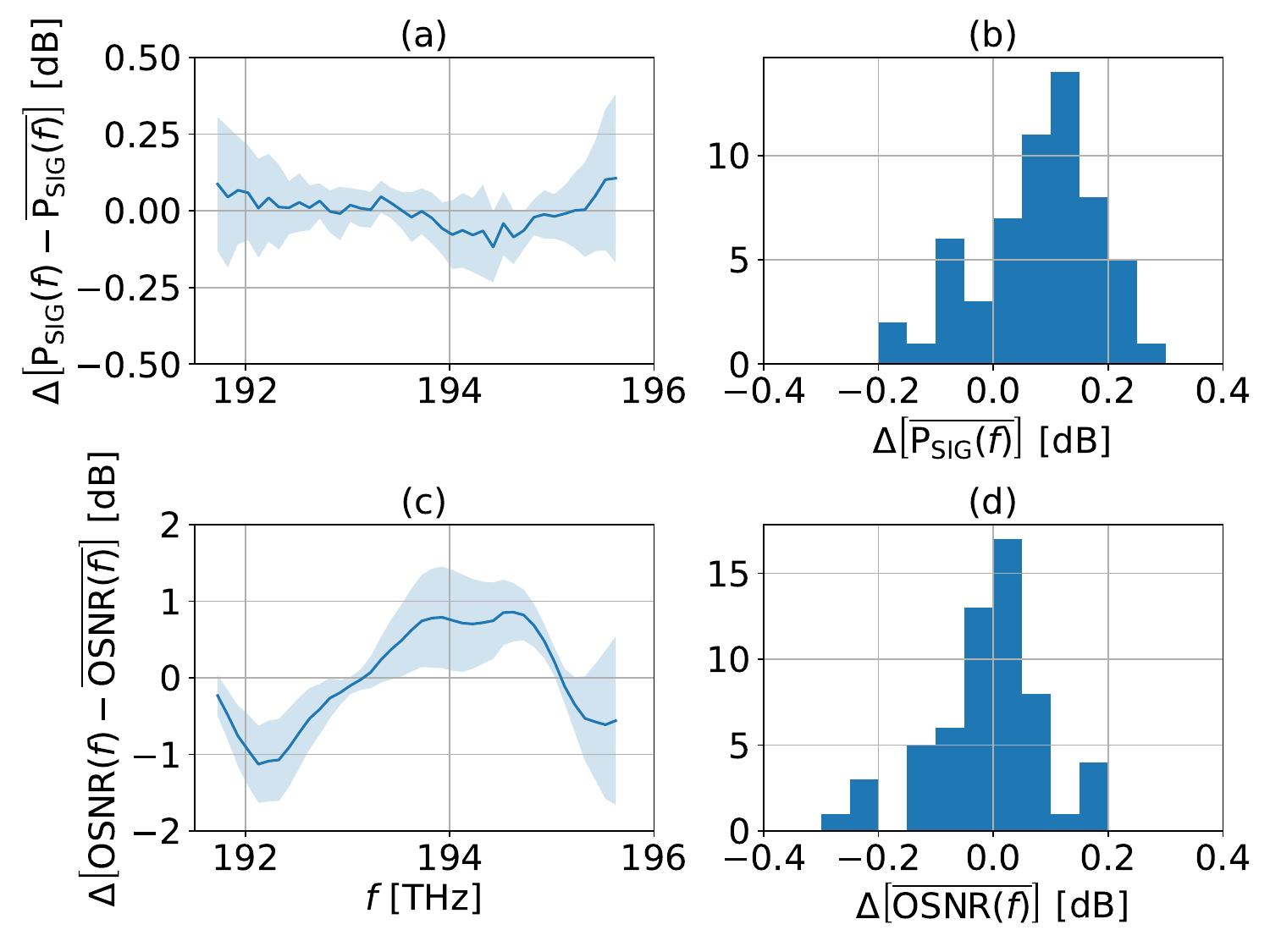}}
\caption{Signal power and OSNR accuracy achieved on the collected dataset: (a) signal power ripple error; (b) signal power average error; (c) OSNR ripple error; (d) OSNR average error.}
\label{fig:calibration_accuracy}
\end{figure}

\begin{figure*}[!t]
\centerline{\includegraphics[width=\linewidth]{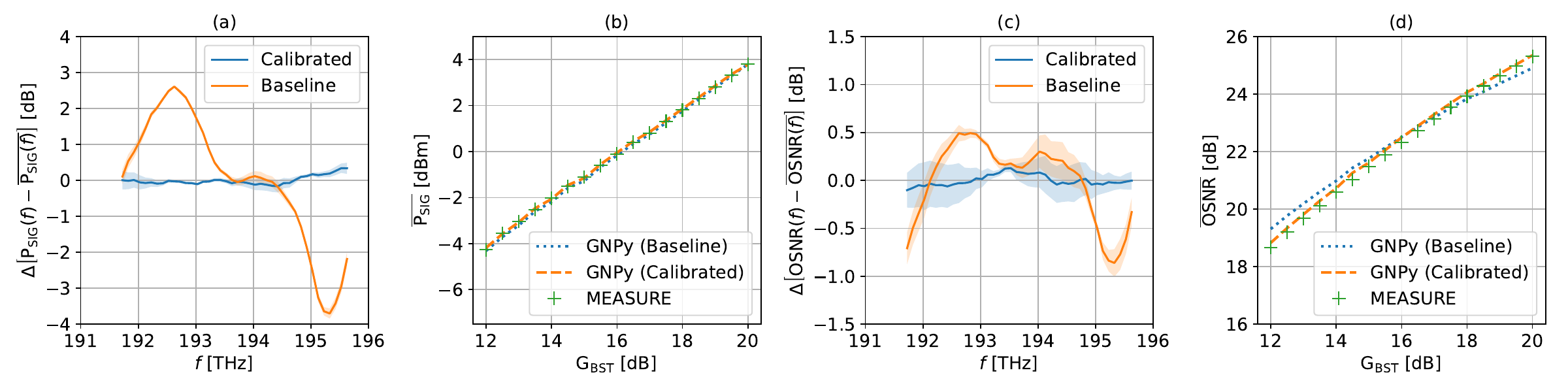}}
\caption{Power sweep in transparency configuration: carrier link validation and baseline comparison. (a) signal power ripple error; (b) signal power average; (c) OSNR ripple error; (d) OSNR average.}
\label{fig:carrier_link_psig_osnr}
\end{figure*}

Feeding GNPy with the retrieved set of physical parameters, Fig.~\ref{fig:calibration_accuracy} shows the achieved signal power, $\mathrm{P_{SIG}}$, and $OSNR$ accuracy downstream of the calibration procedure, counting 58~measurements of different gain and tilt setting EDFA configurations.
The extracted physical parameters allow the model to achieve an excellent prediction regarding the average, $\overline{\mathrm{P_{SIG}}(f)}$, and the ripple, $\mathrm{P_{SIG}}(f) - \overline{\mathrm{P_{SIG}}(f)}$, of the signal power (although the latter has a slight bias of 0.1~dB), and the OSNR average, $\overline{\mathrm{OSNR}(f)}$, respectively in Figs.~\ref{fig:calibration_accuracy}-a, \ref{fig:calibration_accuracy}-b and \ref{fig:calibration_accuracy}-d. 
On the other hand, the OSNR ripple, $\mathrm{OSNR}(f) - \overline{\mathrm{OSNR}(f)}$, has a variable error in frequency with fluctuations in specific areas denoted by the larger standard deviation (shaded area of Fig.~\ref{fig:calibration_accuracy}-c), $3\,\sigma_{\Delta[\mathrm{OSNR}(f) - \overline{\mathrm{OSNR}(f)}]}$.
This is mainly due to the fact that the model does not consider the frequency variation of the noise figure.
However, the outcome of the proposed methodology provides important information in this regard, allowing us to make a correction on the OSNR in frequency by exploiting the average error profile on the ripple and defining a margin regarding the fluctuation of this ripple.
Since the application of this procedure is disaggregated by optical lines, this allows to increase the accuracy of the QoT estimate.

\begin{table}[!b]
\centering
\begin{tabular}{c|c|c|c|c|c|c|}
\cline{2-7}
 &
  \begin{tabular}[c]{@{}c@{}}AAL 1\\ PRE\end{tabular} &
  \begin{tabular}[c]{@{}c@{}}CL\\ BST\end{tabular} &
  \begin{tabular}[c]{@{}c@{}}CL\\ ILA 1\end{tabular} &
  \begin{tabular}[c]{@{}c@{}}CL\\ ILA 2\end{tabular} &
  \begin{tabular}[c]{@{}c@{}}CL\\ PRE\end{tabular} &
  \begin{tabular}[c]{@{}c@{}}AAL 2\\ BST\end{tabular} \\ \hline
\multicolumn{1}{|c|}{\begin{tabular}[c]{@{}c@{}}$\mathrm{G}$ $[$dB$]$\end{tabular}}                     & 21.1 & 19.4 & 15.5 & 14.9 & 16.8 & 16.4 \\ \hline
\multicolumn{1}{|c|}{\begin{tabular}[c]{@{}c@{}}$\mathrm{T}$ $[$dB$]$\end{tabular}}                     & 0.0  & -1.3 & -1.3 & -1.3 & -1.3 & -1.5 \\ \hline
\multicolumn{1}{|c|}{\begin{tabular}[c]{@{}c@{}}$\mathrm{P_{OUT}}$\\ $[$dBm$]$\end{tabular}} & 13.0 & 19.3 & 19.5 & 19.5 & 19.5 & 12.0 \\ \hline
\end{tabular}
\caption{EDFA telemetry data in transparency configuration.}
\label{tab:transparency_config}
\end{table}

\subsection{Performance Validation \& Baseline Comparison}

The model calibrated using the physical parameters retrieved has been validated performing a power sweep of the carrier link booster amplifier and compared in terms of accuracy with a baseline model.

The carrier link baseline model has been built using standard values for SSMF and a fixed noise figure of 7~dB for EDFAs.
The input and output connector losses are assumed equal to 1.5 and 0.5~dB, respectively.
The loss coefficient is assumed flat along the C-band and it is estimated diving the total loss of the considered fiber span measured by the EDFA photodiodes in the transparency configuration for the fiber span length.
In the computation, the value of measured total loss has been purged by the assumed connector losses.

The transparency configuration has been calculated using the calibrated model where each EDFA recovers the loss of the previous span, the booster is set in order to maximize the Q-factor, and the tilt of each EDFA is set at a single value in order to minimize the tilt of the signal power spectrum at the receiving edge node.
The EDFA telemetry information in the defined transparency configuration is reported in Tab.~\ref{tab:transparency_config}.
The EDFAs belonging to the AALs are set in constant output power at a value allowing spectrum equalization at the first carrier link edge node and sufficient power budget at the receiver.

\begin{figure}[!b]
\centerline{\includegraphics[width=\linewidth]{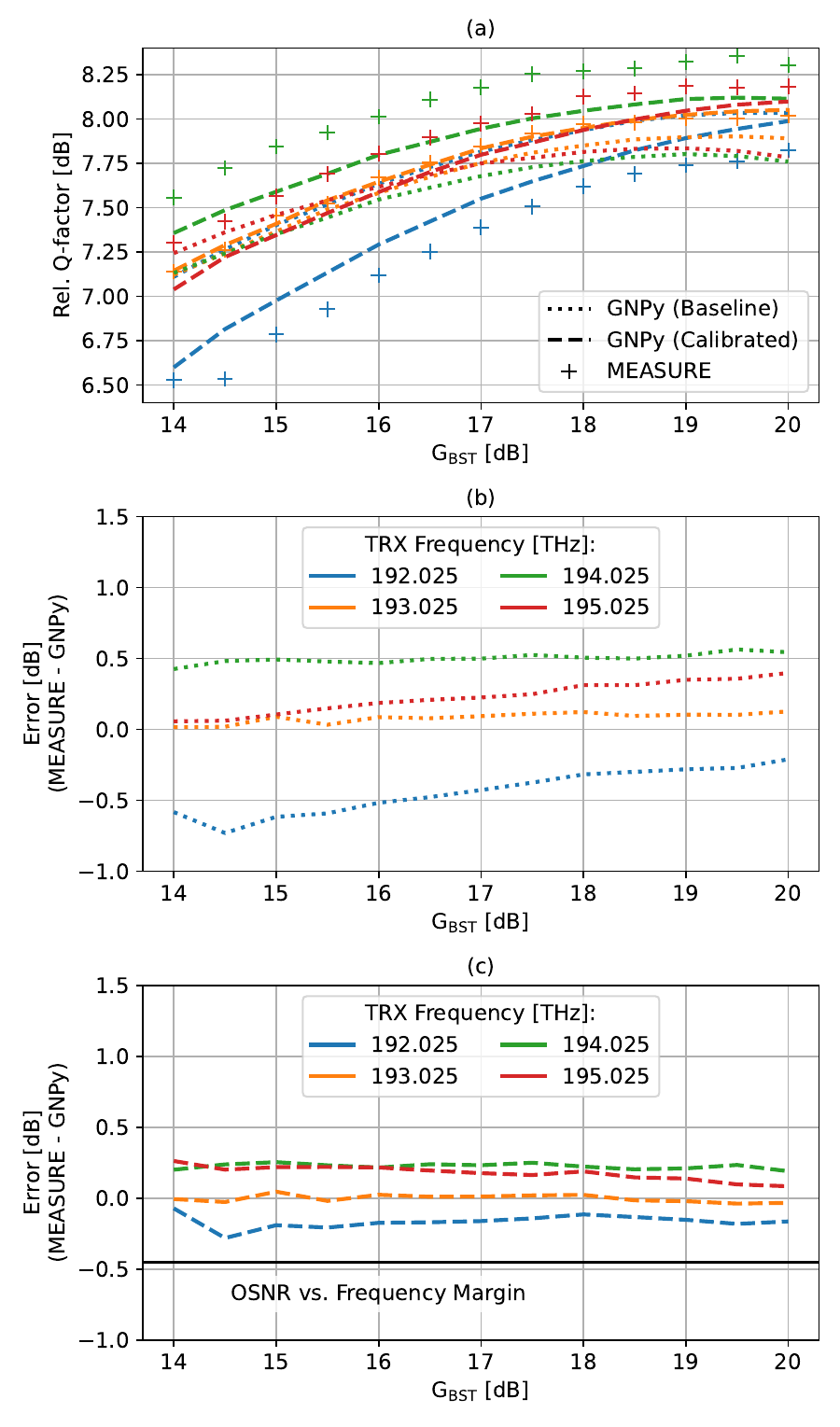}}
\caption{Power sweep in transparency configuration: validation and baseline comparison. (a) relative Q-factor; (b) relative Q-factor error using the baseline model; (c) relative Q-factor error using the calibrated model.}
\label{fig:validation_transparency_config_dlm_calib}
\end{figure}

The power sweep is performed progressively increasing the carrier link booster from 12 to 20~dB with a step of 0.5~dB and maintaining the other EDFAs in transparency configuration.
Fig.~\ref{fig:carrier_link_psig_osnr} shows the results related to the validation and baseline comparison of the carrier link in terms of signal power and OSNR prediction.
In particular, Figs.~\ref{fig:carrier_link_psig_osnr}-a and \ref{fig:carrier_link_psig_osnr}-c represent the distribution over frequency in terms of ripple error with respect to the measurements of the two metrics considered, and Figs.~\ref{fig:carrier_link_psig_osnr}-b and \ref{fig:carrier_link_psig_osnr}-d report the average measured values and the related predictions of the two models.
It is evident that the model that exploits the parameters extracted with the proposed calibration outperforms the baseline model both in terms of signal power and OSNR estimation.
In particular, with regards to signal power, it is important to note how modeling the ripple of the amplifiers and the attenuation of the fibers in frequency allows an adequate estimate of the power budget at the receiving node.
This impacts the transmission strategy downstream of the line, such as the equalization achieved by the ROADMs.
Regarding the OSNR, firstly the calibrated model allows to follow the average as the booster gain increases accurately, compared to the baseline model which considers a fixed noise figure.
Secondly, the information coming from the measurements of the calibration procedure allows to improve the frequency estimate of the OSNR profile despite the calibrated model considering the noise figure flat in frequency.

Fig.~\ref{fig:validation_transparency_config_dlm_calib} shows the validation in terms of relative Q-factor for the EtE connection.
The measured values and the predictions of both models are represented in Fig.~\ref{fig:validation_transparency_config_dlm_calib}-a.
In order to easily compare the model predictions, the relative Q-factor error has been depicted in Fig.~\ref{fig:validation_transparency_config_dlm_calib}-b for the baseline model and in Fig.~\ref{fig:validation_transparency_config_dlm_calib}-c for the calibrated one.
It is worth mentioning that in short-reach transmission scenarios, the dominant contribution that determines the quality of the transmission is the SNR of the transceiver and the factors connected to it, such as filtering penalties.
Afterwards, even if the error curves of the calibrated model are flatter than the baseline model, suggesting a correct estimate of the impairments, the two models do not differ much in the prediction of the relative Q-factor precisely because of the dominant effect of the transceiver.
Another important observation is that the tested field trial does not allow the optimal transmission regime to be observed in detail and does not allow the nonlinear regime to be reached due to the consistent presence of losses concentrated at the entrance of the fiber spans, mainly due to the connections in patch panels.
Focusing on the results of the calibrated model, it is possible to notice that the error curves present some offsets. These residuals can be attributed to any additional impairment of the transceiver, such as variations in frequency of performance or deviations of the back-to-back curve, or a residual deviation of the OSNR in frequency, which is in any case considered as a margin (Fig.~\ref{fig:validation_transparency_config_dlm_calib}-c) .

\subsection{System Q-factor Fluctuation}

\begin{figure}[!t]
\centerline{\includegraphics[width=\linewidth]{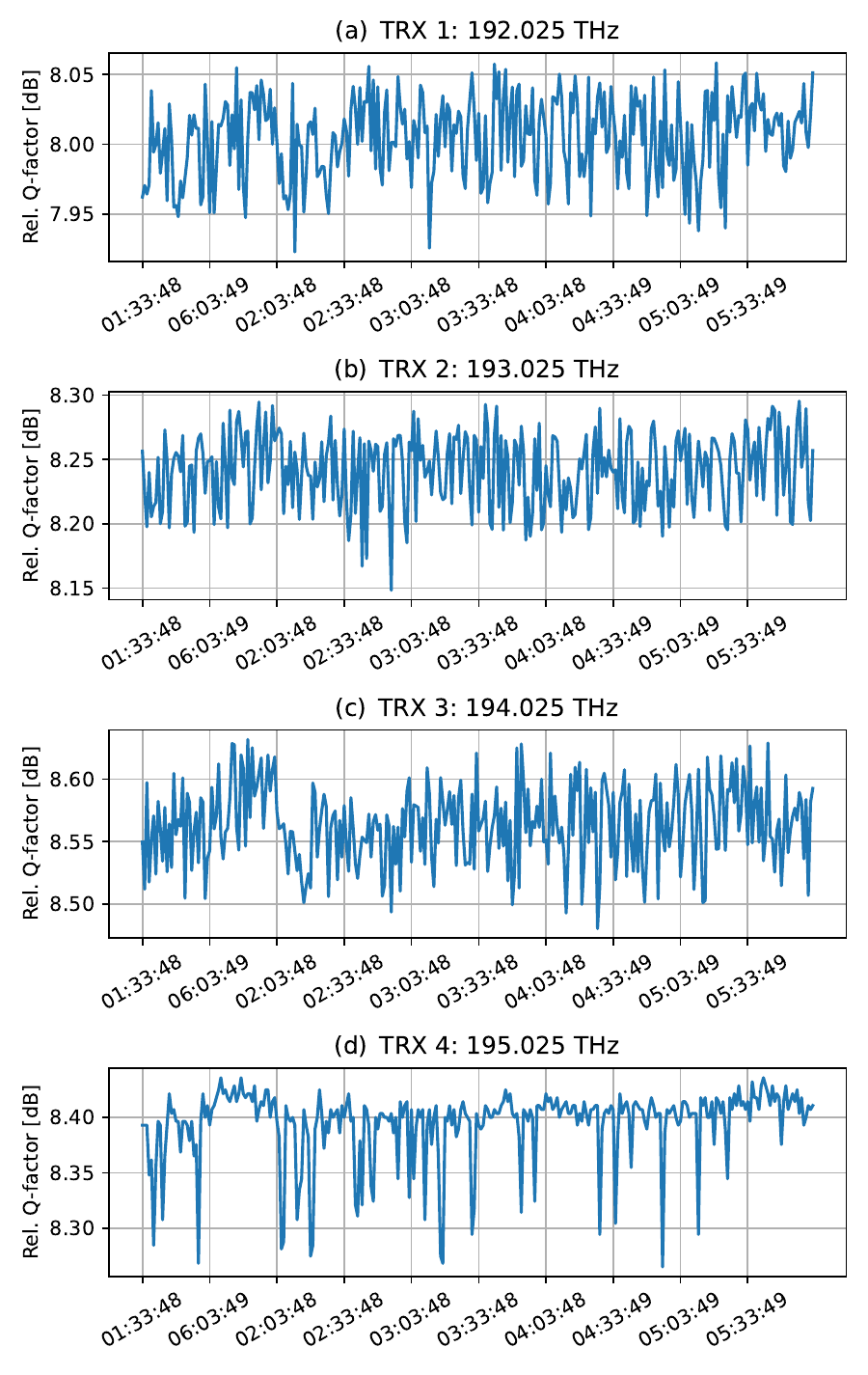}}
\caption{Q-Factor fluctuation over 5~hours time lapse.}
\label{fig:continuos_monitoring_time}
\end{figure}

\begin{figure}[!t]
\centerline{\includegraphics[width=\linewidth]{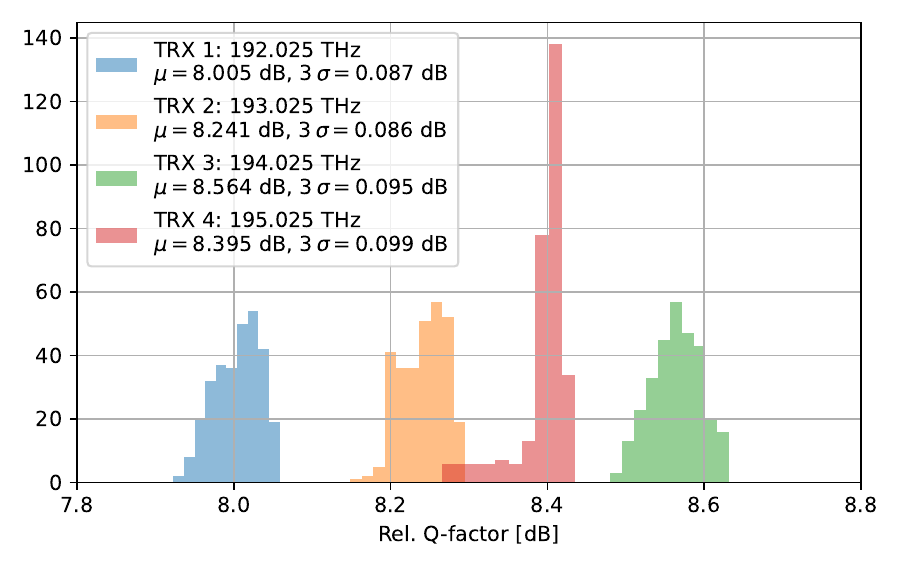}}
\caption{Distribution of Q-Factor data collected in 5~hours.}
\label{fig:continuos_monitoring_distr}
\end{figure}

After optimizing the OLS by using the integrated physical-aware method, we conducted a five-hour quality measurement.
Fig.~\ref{fig:continuos_monitoring_time} shows the five-hour Q-factor fluctuation of the experimental system, and Fig.~\ref{fig:continuos_monitoring_distr} shows the distribution of the Q-factor data. We measured the EtE Q-factor every minute with 4 TRxs.
The standard deviations ($3\,\sigma$) of the four TRx values were 0.087, 0.086, 0.095, and 0.099.
These results indicate that the Q-factor fluctuation was sufficiently small, and that the system was stable even after the semi-automatic fiber-switching work conducted in this experiment.

\subsection{Operational Considerations}

From the experimental results presented in this section, we make the following observations on automatic operation.

\begin{enumerate}[label=\Alph*]
    \item Automation feasibility perspective: Although we used pre-measured lengths for each span, we could retrieve all the necessary characteristic parameters of the optical system with integrated physical-aware methods.
    The visualization function enables visual inspection by human operators, and the semi-automatic sequencing allows reversion in the event of unexpected problems.

    \item Benefits of automation: When telecom operators temporarily interrupt service to perform construction work on a user line that is in service, the maintenance window is typically two hours. With this method, we could switch and calibrate within one-hour via remote operators in a different time zone.
\end{enumerate}

\section{Future Challenges} \label{sec:future_challenges}

In this experiment, we used the Open ROADM MSA standard 400G 16QAM mode in our BER measurements. In DCI, 32QAM and 64QAM with high spectral efficiency are expected to be applied to increase the transmission capacity per fiber. However, these modes require a higher OSNR, and further improvement in the calibration accuracy is thus mandatory. Reduction of the time required for calibration is also an important research issue.
In this paper, a single point-to-point line was optimized, but if the calibration time could be further reduced, larger networks could be constructed within a limited maintenance window. Another important research area involves optimization methods and architectures for scenarios where the network to be constructed contains heterogeneous portions (e.g., some segments have different operational policies, and the control interfaces of OLS devices are not disclosed).
In addition, the 1-hour remote maintenance and provisioning shown in Chapter 3 was tested in a field environment, while the multi-vendor device control using the hybrid controller shown in Chapter 4 was conducted only in a laboratory environment. If all of these operations could be controlled automatically from a remote location, this would be a good practical example of the digital twin in optical transport.
As for the user support viewpoint, in this experiment we only focused on OLS provisioning as an operational use case, but fault detection is another attractive use case for this technology.
Finally, in this experiment we implemented the DLM functionality in an embedded transponder rather than a pluggable module for data communication. In the future, DLM technology should be made available for implementation in a wide range of coherent modules.

\section{Conclusion} \label{sec:conclusion}

We proposed an approach and architecture to semi-automatically measure and optimize newly installed fiber optic line systems using integrated physical-parameter-aware technologies. We focus on how our approach and key enablers can assist operational automation of optical networks and solve commercialization issues. For the optimization method, we started and demonstrated real-time integration of DLM and OL Calibration to retrieve physical parameters for OLS performance optimization. In paticular, we successfully demonstrated semi-automatic OLS provisioning on field fiber networks at the Duke University testbed in Durham, North Carolina. By applying a hybrid controller architecture with OSS, we verified that construction work at a local site could be conducted from a remote site in a different time zone.
Our optimization method has the following benefits: minimum footprint at the user site, accurate estimation of the critical network characteristics through complementary telemetry techniques, and the capability for all operational tasks to be performed remotely. The integrated physical-parameter-aware technique enables acquisition of all the necessary parameters of optical system and provide visualization for human inspection. The semi-automatic control sequencing is designed such that the operation can reverted in the event of unexpected problems. Through this approach, we have shown that a remote operator in a different time zone can perform switchover and calibration in less than one hour. By comparing the measured Q-factor with the estimates calculated by GNPy, we showed that our physical-parameter-aware physical model can improve the QoT prediction error over existing designs.

\section*{Funding} These research results were obtained with the support of a grant program (adoption number 50201) by the National Institute of Information and Communications Technology (NICT), Japan. This work was also supported in part by NSF grants CNS-2211944 and CNS-2330333.

\section*{Acknowledgments} We thank the Telecom Infra Project OOPT-PSE working group and the Duke Office of Information Technology (OIT) team for their support.

\bibliography{references}
  
\end{document}